\documentclass[prb,twocolumn,showpacs,preprintnumbers,amsmath,amssymb]{revtex4}

\usepackage{color}
\usepackage{graphicx}
\usepackage{dcolumn}
\usepackage{bm}

\preprint{submitted to the Physical Review B}

\begin{document}

\title{Ab-initio determined
electronic and magnetic properties of half-metallic NiCrSi and
NiMnSi Heusler alloys; the role of interfaces and defects}

\author{I. Galanakis}\email{galanakis@upatras.gr}
\affiliation{Department of Materials Science, School of Natural
Sciences, University of Patras,  GR-26504 Patra, Greece}

\author{K. \"Ozdo\u{g}an}\email{kozdogan@gyte.edu.tr}
\affiliation{Department of Physics, Gebze Institute of Technology,
Gebze, 41400, Kocaeli, Turkey}

\author{E. \c{S}a\c{s}{\i}o\u{g}lu}\email{e.sasioglu@fz-juelich.de} \affiliation{Institut f\"ur
Festk\"orperforschung,
Forschungszentrum J\"ulich, D-52425 J\"ulich, Germany\\
Fatih University,  Physics Department, 34500,    B\" uy\" uk\c
cekmece,  \.{I}stanbul, Turkey}

\begin{abstract}
Using state-of-the-art first-principles calculations we study the
properties of the ferromagnetic Heusler compounds NiYSi where Y
stands for V, Cr or Mn. NiCrSi and NiMnSi contrary to NiVSi are
half-metallic at their equilibrium lattice constant exhibiting
integer values of the total spin magnetic moment and thus we
concentrate on these two alloys. The minority-spin gap has the
same characteristics as for the well-known NiMnSb alloy being
around $\sim$1 eV. Upon tetragonalization the gap is present in
the density of states even for expansion or contraction of the
out-of-plane lattice parameter by 5\%. The Cr-Cr and Mn-Mn
interactions make ferromagnetism extremely stable and the Curie
temperature exceeds 1000 K for NiMnSi. Surface and interfaces with
GaP, ZnS and Si semiconductors  are not half-metallic but in the
case of NiCrSi the Ni-based contacts present spin-polarization at
the Fermi level over 90\%. Finally, we show that there are two
cases of defects and atomic-swaps. The first-ones which involve
the Cr(Mn) and Si atoms induce states at the edges of the gap
which persists for a moderate-concentration of defects. Defects
involving Ni atoms induce states localized within the gap
completely destroying the half-metallicity. Based on
single-impurity calculations we associate these states to the
symmetry of the crystal.
\end{abstract}


\maketitle

\section{Introduction\label{sec1}}

Magnetoelectronics, also known as spintronics, is probably the
field with the most rapid growth in materials science bringing to
the center of scientific research new
phenomena.\cite{Zutic,Felser,Zabel} For researchers dealing with
ab-initio calculations one of the most interesting concept in
spintronics is the half-metallicity.\cite{book,Review1,Review2}
Research has been focused initially on the explanation of the
origin of the half-metallicity and afterwards on the prediction of
new half-metallic materials with potential applications in
spintronics devices. Half-metals are hybrids between normal metals
and semiconductors: the majority-spin band is crossed by the Fermi
level as in a normal metal while the Fermi level falls within a
gap in the minority-spin band as in semiconductors leading to a
perfect 100\%\ spin-polarization at the Fermi level,\cite{Review1}
maximizing the efficiency of magnetoelectronic devices.\cite{Wolf}
de Groot and his collaborators in 1983 were the first to predict
the existence of half-metallicity in the case of the intermetallic
semi-Heusler alloy NiMnSb,\cite{deGroot} and the origin of the gap
seems to be well understood.
\cite{GalaHalf,Jung,Nanda1,Nanda2,Offernes07,Kohler} There exist
several ab-initio calculations on NiMnSb reproducing the initial
results of de Groot and collaborators,\cite{calc1,calc2,calc3} and
Galanakis \textit{et al.} showed that the gap arises from the
hybridization between the $d$ orbitals of the Ni and Mn
atoms.\cite{GalaHalf} Moreover in the latter reference it was
shown that the total spin moment for the Heusler alloys of the
NiMnSb type follows a Slater-Pauling behavior being in $\mu_B$ the
total number of valence electrons in the unit cell minus 18 since
there are exactly nine occupied minority-spin
states.\cite{GalaHalf} Theoretical calculations for the interfaces
of these materials with the semiconductors are few and all results
agree that in general the half-metallicity is lost both at the
surfaces\cite{GalaSurf1,GalaSurf2,GalaSurf3,Jenkins} and the
interfaces with binary
semiconductors.\cite{groot2,GalaInter,Debern,LezaicSTM} Wijs and
de Groot have argued than in the case of the NiMnSb/CdS (111)
contacts the Sb/S interface keeps half-metallicity when the S
atoms sit exactly on top of Sb.\cite{groot2} Moreover taking into
account also the reconstruction at the surfaces and interfaces can
have an important effect on their properties.\cite{Hashefimar}
Finally we should note that several other aspects of NiMnSb have
been studied using first-principles calculations like the exchange
constants and Curie
temperatures,\cite{Sasioglu,SasiogluAPL,SasiogluPRB} the
quasiparticle states\cite{Chioncel} and the dynamical
effects,\cite{Chioncel06} the defects and
doping,\cite{Orgassa,Alling,Attema}
 the structural stability,\cite{Larson,Block} the effect of spin-orbit
 coupling,\cite{Mavropoulos} the fit of model Hamiltonians to
 describe NiMnSb,\cite{Sandratskii,Yamasaki} orbital magnetism,\cite{GalaOrbit} the pressure
 and thermal expansion effects,\cite{Pugaczowa} the temperature effect\cite{Attema07} and the
 magneto-optical properties.\cite{Antonov,GalaXMCD}

The half-metallic character of NiMnSb in single crystals has been
well-established experimentally. Infrared
absorption\cite{Kirillova95} and spin-polarized
positron-annihilation\cite{Hanssen1,Hanssen2} gave a
spin-polarization of $\sim$100\% at the Fermi level. High quality
films of NiMnSb have been
grown,\cite{Roy1,Roy2,Roy3,Giapintzakis1,Giapintzakis2,Giapintzakis3}
 but they were found not to reproduce the half-metallic character of
the bulk. Values of 58\% and 50\% for the spin-polarization at the
Fermi level were obtained by Soulen \textit{et al.}\cite{Soulen98}
and by Mancoff \textit{et al.},\cite{Mancoff99} respectively,  and
recently Zhu \textit{et al.}\cite{Zhu01} found a value of 40\%
using spin-resolved photoemission measurements on polycrystalline
films. Ristoiu \textit{et
al.}\cite{Ristoiu1,Ristoiu2,Ristoiu3,Ristoiu4} showed that during
the growth of the NiMnSb thin films, Sb atoms segregate to the
surface decreasing the obtained spin-polarization; they measured a
value of $\sim$30\% at 200K, while at room temperature the net
polarization was practically zero. But when they removed the
excess of Sb by a flash annealing, they managed to get a nearly
stoichiometric ordered alloy surface terminating in MnSb. Inverse
photoemission experiments at room temperature revealed that the
latter surface shows a spin-polarization of about 67$\pm$9\% which
is significantly higher than all previous values.\cite{Ristoiu2}
There is also experimental evidence that for a temperature of
$\sim$80 K there is transition from a half metal to a normal
ferromagnet,\cite{Hordequin1,Hordequin2} but these experiments are
not yet conclusive. Finally, the effect of stress on the magnetic
anisotropy of thin NiMnSb films and the role of defects have been
explored.\cite{botters,nowicki}

 Based on the success of first-principles electronic
structure calculations to describe the properties of NiMnSb,
several authors have predicted new half-metallic Heusler alloys
crystallizing in the $C1_b$ structure of semi-Heusler compounds
like NiCrM and NiVM (M= P, As, Sb, S, Se and
Te),\cite{Zhang2003,Zhang2004,Sasioglu2005} and XCrAl (X= Fe, Co,
Ni) and NiCrZ (Z= Al, Ga, In).\cite{Luo08} Recently,
Katayama-Yoshida and collaborators published a paper including
also ab-initio calculations on NiMnSi semi-Heusler alloy,
 which was predicted to have a Curie temperature of 1050 K,\cite{Katayama}
 exceeding even the 730 K shown by NiMnSb.\cite{Landolt}
 This finding motivated us to study the electronic and magnetic
properties of this compound in detail since except the very high
Curie temperature, it should be easily grown due to the well-known
and well-controlled growth of NiMnSb and it should be compatible
with Si (Si crystallizes in a diamond structure and the unit cell
of NiMnSi is two times the unit cell of Si). We decided to expand
our study to cover also the closely-related NiCrSi and NiVSi
alloys. NiVSi was found not to be half-metallic at its equilibrium
lattice constant and thus we focused our study on NiCrSi and
NiMnSi alloys. In section \ref{sec2} we present the details of our
calculations and in section \ref{sec3} the electronic and magnetic
properties of the bulk phase of these alloys. In section
\ref{sec4} we discuss the robustness of half-metallicity and
ferromagnetism, and in section \ref{sec5} the properties of (100)
surfaces and of their interfaces with the Si, GaP and ZnS
semiconductors which have an experimental lattice constant almost
identical to the theoretical equilibrium lattice constants of
NiCrSi and NiMnSi. Finally in section \ref{sec6} we discuss the
effect of defects and disorder and in section \ref{sec7} we
summarize and conclude.

\begin{figure}
\includegraphics[scale=0.28]{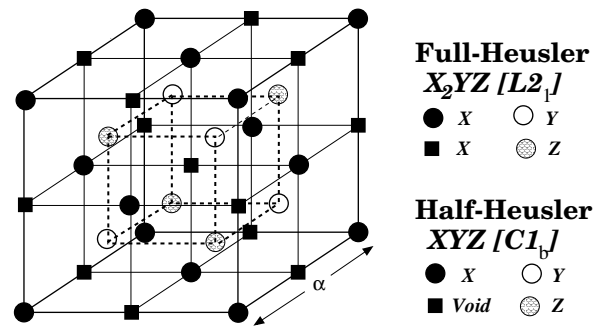}
 \caption{\label{fig1} Schematic representation of the $C1_b$ structure adopted by
 the semi-(or half-)Heusler compounds like NiMnSi and of the
 $L2_1$ structure adopted by the
 full-Heusler compounds like Co$_2$MnSi. The lattice
consists of 4 fcc sublattices. The unit cell in the case of
semi-Heuslers XYZ is that of a fcc lattice with four atoms as
basis: X at $(0\:0\:0)$, Y at $({1\over4}\:{1\over4}\:{1\over4})$,
a vacant site at $({1\over2}\:{1\over2}\:{1\over2})$  and Z at
$({3\over4}\:{3\over4}\:{3\over4})$  in Wyckoff coordinates. In
the case of full-Heusler compounds the vacant site is also
occupied by a X atom.}
\end{figure}

\begin{figure}
\centering
\includegraphics[scale=0.43]{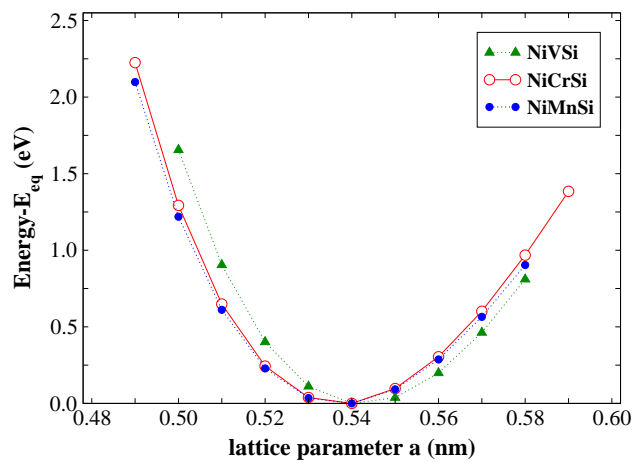}
\caption{Calculated total energy within the FPLO method vs the
lattice constant. All three compounds have a theoretical
equilibrium lattice constant close to 0.54 nm. Note that we have
scaled for each compound the total energy so that the zero energy
corresponds to the one of the equilibrium lattice constant.}
\label{fig2}
\end{figure}

\section{Description of present calculations\label{sec2}}

To study the NiYSi alloys we have employed both the
full--potential nonorthogonal local--orbital minimum--basis band
structure scheme (FPLO)\cite{koepernik1,koepernik2} and the
full-potential screened Korringa-Kohn-Rostoker (FSKKR)
method.\cite{Pap02,Zeller08} Both methods as we will discuss later
give similar results for the bulk alloys. FPLO is suitable to
study the disorder within the coherent-potential approximation
(CPA). FSKKR is suitable to study surfaces and interfaces since
its screened character and the use of decimation method for the
inversion of the Green function matrix lead to an order $N$
scaling of the CPU time, where $N$ is the number of inequivalent
atoms in the unit cell. Moreover Dyson equation makes possible the
study of single impurities within the framework of FSKKR in real
space.\cite{imp1,imp2} In both cases we employed the
local-spin-density approximation (LSDA)\cite{LDA} for the
exchange-correlation energy within the framework of the density
functional theory.\cite{kohn1,kohn2}

\begin{figure}
\centering
\includegraphics[scale=0.43]{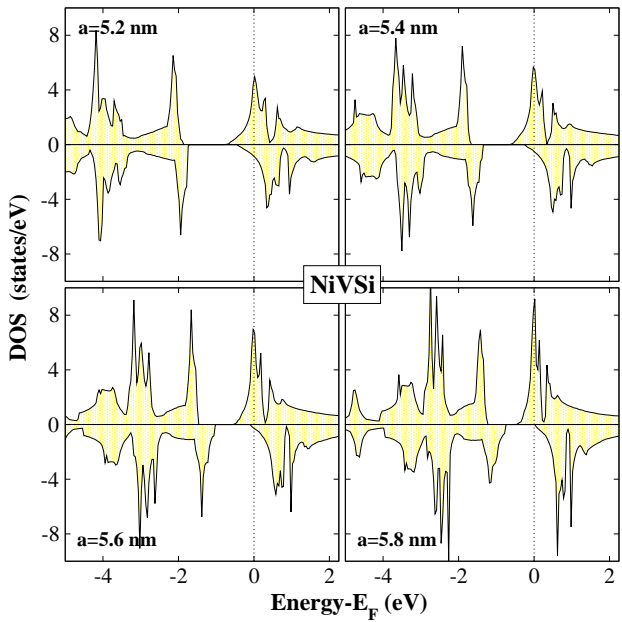}
\caption{ (Color online)FPLO calculated total density of states
(DOS) for the NiVSi compound for four different values of the
lattice constant. Positive values of the DOS correspond to the
majority (spin-up) electrons and negative values to the minority
(spin-down) electrons. The zero of the energy has been chosen as
the Fermi energy.} \label{fig3}
\end{figure}

Semi-Heusler alloys (also known as half-Heusler compounds)
crystallize in the $C1_b$ structure which consists of four fcc
sublattices and have the chemical formula XYZ  where X a
high-valent transition metal atom, Y a low-valent transition metal
atom and Z a $sp$ atom. In Fig. \ref{fig1} we present both the
$C1_b$ structure of semi-Heuslers and the $L2_1$ structure of the
full-Heusler compounds like Co$_2$MnSi. The unit cell is that of a
fcc lattice with four atoms as basis: X atoms occupy the
$(0\:0\:0)$ site, Y atoms the $({1\over4}\:{1\over4}\:{1\over4})$,
the $({1\over2}\:{1\over2}\:{1\over2})$ site is vacant for the
semi-Heuslers and occupied by X in the case of full-Heuslers, and
finally Z atoms occupy the $({3\over4}\:{3\over4}\:{3\over4})$
site (we use Wyckoff coordinates to denote the sites).  The
zinc-blende structure adopted by a large number of semiconductors,
like GaP and ZnS is also consisting of four fcc sublattices. In
the case of GaP the $(0\:0\:0)$ site is occupied by a Ga atom, the
$({1\over4}\:{1\over4}\:{1\over4})$ site by a P atom, while the
other two sites are unoccupied. The diamond structure of Si is
also coherent with the $C1_b$ structure of semi-Heusler alloys
since double the unit cell gives the zinc-blende structure of
binary semiconductors. This close structure similarity makes the
Heusler alloys compatible with the existing semiconductor
technology and thus very attractive for industrial applications.

Since the compounds under study have not been grown
experimentally, we have to determine their equilibrium lattice
constant. In Fig. \ref{fig2} we plot the FPLO calculated total
energies vs the lattice constant for all three NiVSi, NiCrSi and
NiMnSi. We have assumed that all three are ferromagnets (which is
confirmed by our calculated exchange constants, at least for
NiCrSi and NiMnSi, in section \ref{sec4}). We have scaled the
total energy for each compound in a way that the zero of the total
energy corresponds to the minimum energy at the theoretical
equilibrium lattice constants. All three compounds have an
equilibrium lattice constant very close to 5.4 \AA . We have
decided to use the latter one for all calculations presented in
this manuscript.

\begin{table}
\centering \caption{ (Color online)Atom-resolved and total spin
magnetic moments in $\mu_B$ calculated using both the FPLO and
FSKKR methods. In the last two columns we present our results on
the Curie temperature in Kelvin units calculated with both the
mean-field ($T_C^{MFA}$) and random-phase ($T_C^{RPA}$)
approximations. \label{table1}}
\begin{tabular}{l|c|c|c|c|c|c} \hline

Compound & $m^{Ni}$ & $m^{Y}$ & $m^{Si}$ & $m^{total}$ & $T_C^{MFA}$ & $T_C^{RPA}$\\

NiVSi (FPLO)& 0.094 & 0.791 & -0.011 & 0.874 & 201 & 174 \\

NiCrSi (FPLO) & 0.127 & 2.020 & -0.148 & 2.000 & 900 & 698 \\

NiCrSi (FSKKR) & 0.167 & 1.858 & -0.089  & 1.969 &&\\

NiMnSi (FPLO) & 0.207 & 3.005  & -0.212 & 3.000 & 1505 & 1120 \\

NiMnSi (FSKKR) & 0.231 & 2.845  & -0.134 & 2.965 && \\ \hline

\end{tabular}
\end{table}

\begin{figure}
\centering
\includegraphics[scale=0.43]{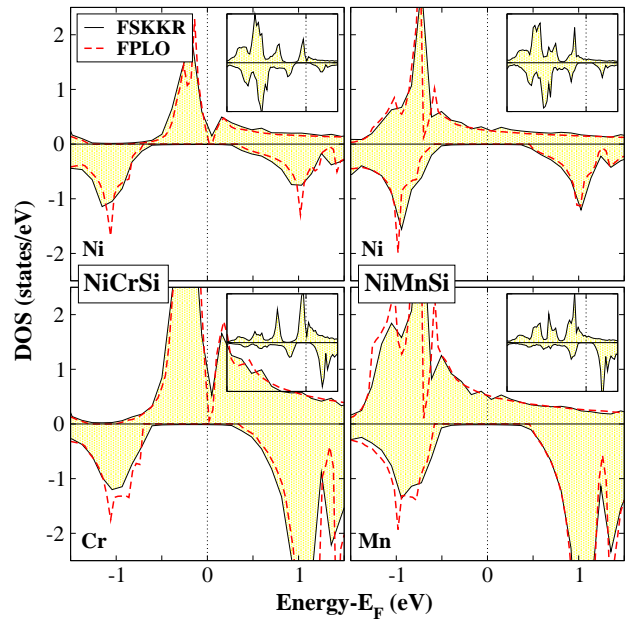}
\caption{(Color online)Ni and Cr(Mn) resolved DOS for both NiCrSi
and NiMnSi compounds using both the FPLO and FSKKR methods. In the
insets we present the DOS for a wider energy range. } \label{fig4}
\end{figure}

We should also discuss here the reason that led us to neglect
NiVSi from our study. In Fig. \ref{fig3} we present the total DOS
for NiVSi and for four different lattice constants: 5.2, 5.4, 5.6
and 5.8 nm (we remind here that 1 nm = 10 \AA ). There is a gap
both in the majority and minority spin bands and the Fermi level
for the equilibrium lattice constant crosses the conduction
minority spin-band and the alloy is not half-metallic. Moreover it
is located in a peak of the majority-spin DOS making
ferromagnetism unstable and probably the compound prefers to be
non-magnetic. The calculated total spin moment presented in Table
\ref{table1} is 0.874 $\mu_B$ lower than the ideal 1.0 $\mu_B$
requested by the Slater-Pauling behavior for half-metallic
semi-Heusler alloys.\cite{GalaHalf} We have not calculated the
difference between the non-magnetic and ferromagnetic states since
it exceeds the scope of this paper. Although the Fermi shifts
lower in energy as we expand the lattice and the compound becomes
half-metallic the extremely high majority DOS at the Fermi level
persists and ferromagnetism seems to be unstable. The latter
argument is also confirmed by the calculated exchange constants
presented in Fig. \ref{fig7}. The Ni-V interaction is negligible
and the V-V exchange constant between vanadium nearest-neighbors
presented in the figure is very weak being 8 times smaller than
the Mn-Mn interaction in NiMnSi and thus the tendency to
ferromagnetism is not very strong. Also the Curie temperature
deduced using these exchange constants is $\sim$200 K within both
the mean-field and random-phase approximations and thus it is
considerably smaller than the room temperature at which realistic
devices should operate. Since NiVSi is not so interesting for
applications with respect to NiCrSi and NiMnSi, we have decided
not to include it in the rest of our study.

\begin{figure}
\centering
\includegraphics[width=\linewidth]{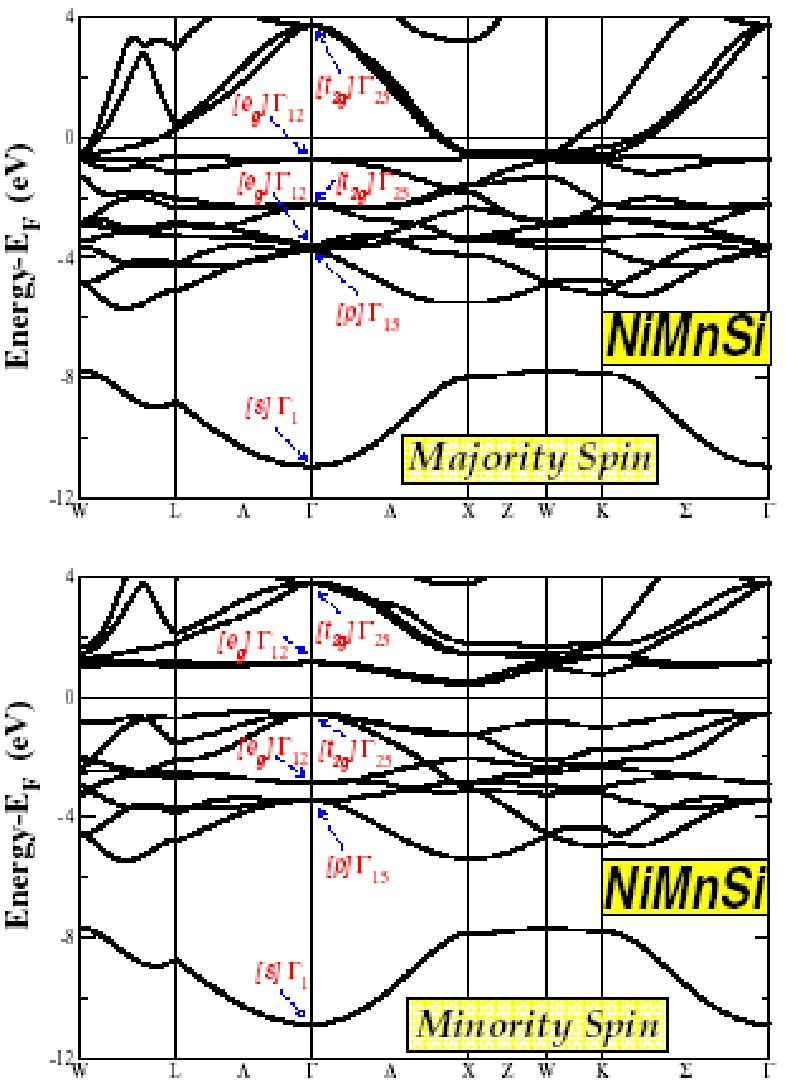}
\caption{ (Color online)FSKKR-calculated spin-resolved band
structure for the NiMnSi compound along several high-symmetry
axis. We have denoted the character of the bands at the $\Gamma$
point and in the brackets we denote the orbitals transforming
following each representation.} \label{fig5}
\end{figure}

\section{Electronic and magnetic properties \label{sec3}}

We will start our discussion from the electronic and magnetic
properties of NiCrSi and NiMnSi establishing also the equivalence
between the FPLO and FSKKR methods. In Fig. \ref{fig4} we present
the DOS of the transition-metal atoms Ni and Cr(Mn) since the DOS
of Si is negligible within both FPLO (dashed line) and FSKKR
(solid line). Both methods give almost the same DOS and the only
difference is the intensity of some peaks. Especially concerning
the gap (in the right panel we have blown-up the region around the
Fermi level) both FPLO and FSKKR give the same gap-width, the same
position of the Fermi level with respect to the edges of the gap
and the same majority-spin DOS at the Fermi level. In Table
\ref{table1} we present also the spin moments using both
electronic structure methods. FPLO gives integer values of the
total spin moments while FSKKR gives slightly smaller values due
to the $\ell$ summation used in FSKKR. This problem has been
extensively discussed in Ref. \onlinecite{GalaFull}.  FSKKR
overestimates with respect to FPLO the hybridization between the
Ni and Mn $d$-states. Since FPLO gives a more atomic-like
description of the Cr and Mn, it predicts larger Cr and Mn spin
moments and consequently smaller Ni spin moments to keep constant
the total spin moment. Si spin moment shows also a small variation
between the two methods since its moment arises from the
hybridization with the Ni-Cr(Mn) $d$ hybrids. The total spin
moment does not exactly equals the sum of the atomic spin moments
of the three constituent atoms since also the vacant site carries
a very small spin moment which we do not present in the table.

Since we have established the equilivalence between the FPLO and
FSKKR we will go on with the discussion of the electronic and
magnetic properties of NiCrSi and NiMnSi. NiMnSi has 21 valence
electrons per unit cell, one less than NiMnSb, and NiCrSi 20
valence electrons. In accordance with the Slater-Pauling rule for
the perfect half-metallic semi-Heusler compounds presented in Ref.
\onlinecite{GalaHalf} the calculated total spin moment in the unit
cell is for NiMnSi 21-18= 3 $\mu_B$ and 20-18= 2 $\mu_B$ for
NiCrSi. The spin moment is mainly carried by the low-valent
transition metal atoms Cr and Mn. Ni carries a  small spin moment
of about 0.1-0.2 $\mu_B$ due to the hybridization with the Cr(Mn)
$d$-states. Si shows a small spin moment which is antiparallel to
the moment of the transition-metal atoms since the majority
$p$-states of Si are more extended in energy and thus partially
unoccupied leading to a surplus of minority-spin Si $p$-states and
a negative atomic spin moment.

The origin of the gap is similar to the case of
NiMnSb.\cite{GalaHalf} Ni and Cr(Mn) $d$-states hybridize creating
5 bonding $d$ hybrids and 5 antibonding ones per spin-direction.
Si provides one $s$- and three $p$-states per spin low in energy.
The $p$-states accommodate also $d$ electrons of the transition
metal atoms. In Fig. \ref{fig5} we present the band-structure
along several high symmetry directions for both spin-directions
for NiMnSi. The character of each band at the $\Gamma$ point
reveals the character of the associated orbitals in the real
space. First for the minority-spin band structure there is one $s$
and a triple-degenerated at the $\Gamma$ point $p$-band coming
from the Si atom. Then follow the double-degenerated $e_g$ and
triple-degenerated $t_{2g}$ bands due to the Ni-Mn bonding $d$
hybrids. The $t_{2g}$ band is separated  by an indirect $\Gamma
-X$ gap from the antibonding bands. Thus in total there are
exactly nine occupied minority-spin bands. In the majority spin
band we have 12 instead of 9 atoms. The two out of the three extra
electrons occupy the antibonding $e_g$ states and the last
electron occupies partially the antibonding $t_{2g}$ bands. In the
case of NiCrSi the majority-spin band has to accommodate 11
electrons and not 12 and thus it is enough to occupy the
majority-spin antibonding $e_g$ states. The band associated with
the latter states is very narrow and is also well separated in
energy from the antibonding $t_{2g}$ bands and this is why in the
case of NiCrSi the Fermi level is located in a very narrow deep
separating two peaks  in the Cr majority-spin DOS.

\begin{figure}
\centering
\includegraphics[scale=0.43]{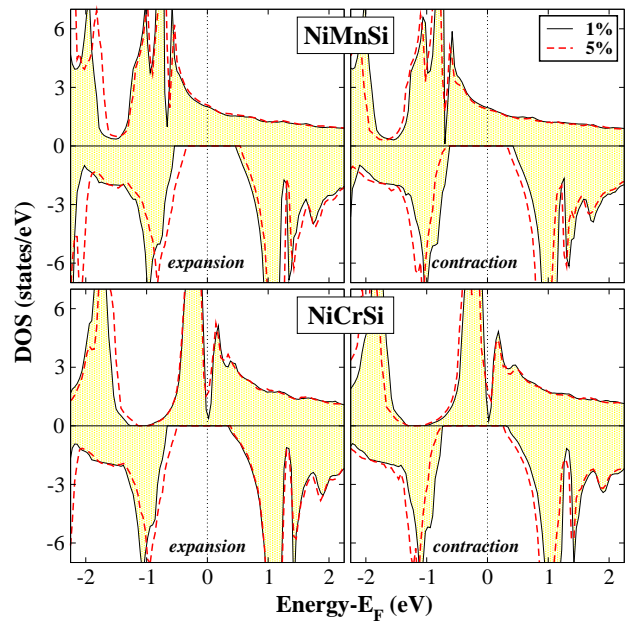}
\caption{ (Color online)FPLO total DOS of NiCrSi and NiMnSi
compounds under 1\%\ and 5\%\ expansion and contraction of the $c$
lattice parameter along the $z$-axis resulting to a
tetragonalization of the lattice. The in-plane lattice parameter
was kept constant and equal to 0.54 nm.} \label{fig6}
\end{figure}

\begin{table}
\centering \caption{ (Color online)FPLO calculated width of the
minority-spin gap in eV for the cases presented in Fig. \ref{fig6}
with respect to the perfect cubic lattice (the ration equals to
zero since $c=a$). Negative values of the ratio in the first
column correspond to contraction and positive values to expansion.
\label{table2}}
\begin{tabular}{r|c|c}\hline
 & \multicolumn{2}{c}{Gap-Width (eV)} \\
$\frac{c-a}{a}$ & NiCrSi & NiMnSi \\

-5\% & 1.029& 0.966 \\

-2\% &  1.067 & 1.050\\

-1\% &  1.071 & 1.055 \\

0\% & 1.071 &  1.055\\

1\% & 1.021 & 1.017\\

2\% & 0.983 & 0.987 \\

5\% & 0.882 & 0.903\\ \hline
\end{tabular}
\end{table}

\section{Robustness of half-metallicity and ferromagnetism \label{sec4}}

We will continue our study discussing the robustness of the
half-metalicity and ferromagnetism. Since alloys grow on top of
semiconductors, the lattice mismatch can lead to a
tetragonalization of the lattice of the Heusler alloys to minimize
the elastic energy of the film. Thus we have to study for
realistic applications the effect of tetragonalization on the
half-metallicity. In Fig. \ref{fig6} we present the total DOS for
NiCrSi and NiMnSi using the FPLO code under contraction or
expansion of the out-of-plane lattice parameter by 1 and 5\%\
keeping the in-plane lattice constants equal to 0.54 nm. The
latter value of tetragonalization is very large and in most
realistic applications the lattice relaxes much less. As can been
deduced from the figure contraction leads to a marginal shift of
the gap while expansion leads to a very small narrowing of the
gap. To give more details on the behavior of the gap we present in
Table \ref{table2} the width of the gap for the different cases.
For the perfect cubic NiCrSi and NiMnSi alloys the gap is 1.071
and 1.055 eV, respectively. Contraction by 5\%\ leads to gap-width
of 1.029 and 0.966 eV respectively, while expansion by 5\%\ to
0.882 and 0.903 eV for NiCrSi and NiMnSi respectively. Thus
tetragonalization leaves the gap almost unchanged and since for
both alloys the Fermi level is near the middle of the gap this has
no effect on the half-metallic properties of the two alloys.

\begin{figure}
\centering
\includegraphics[scale=0.43]{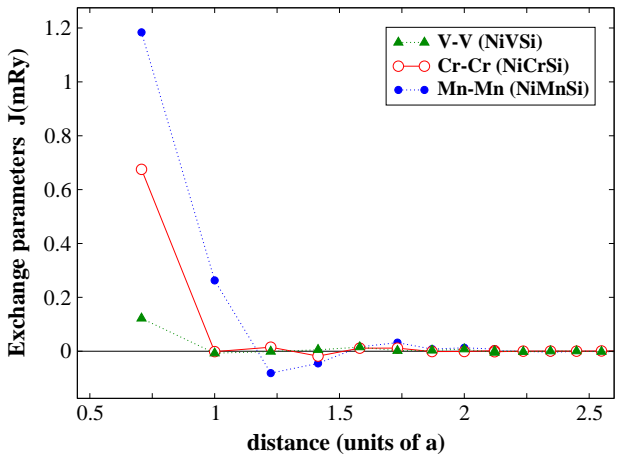}
\caption{ (Color online)Calculated exchange constants between the
low-valent transition metal atoms sitting at the Y site (V, Cr or
Mn) as a function of the distance between them.} \label{fig7}
\end{figure}

Stable half-metallicity does not mean stable ferromagnetism. To
study how robust is the ferromagnetism we have to calculate the
exchange constants and then use them to calculate the Curie
temperature. We have used the formalism presented in Ref.
\onlinecite{Sasioglu} and in Fig. \ref{fig7} we present the Cr-Cr
and Mn-Mn exchange constants as a function of the distance between
the atoms. We neglect the Ni-Cr and Ni-Mn interactions since they
are very small even if each Cr or Mn atom has four Ni atoms as
first neighbors. The much larger Cr and Mn spin moments are
responsible that interactions between them are much larger than
between Ni and Cr(Mn) atoms. In the case of NiMnSi the exchange
constants are very large and even Mn-Mn next-nearest neighbors
contribute to the stability of ferromagnetism. The exchange
constant between nearest neighbors, usually in literature denoted
as $J_1$, is 1.2 mRy. This is a very large value and leads to very
large values for the Curie temperature ($T_C$). Calculating the
latter one as explained in Ref. \onlinecite{Sasioglu} we get a
value of 1505 K within the mean field approximation (MFA) and 1129
K within the random-phase approximation (RPA) (see Table
\ref{table1}) close to the calculated value of 1050 K in Ref.
\onlinecite{Katayama}. Our experience on NiMnSb shows that the RPA
results for semi-Heusler compounds are closer to the experimental
values than MFA. In the case of NiCrSi only the interaction
between Cr atoms nearest-neighbors contributes to the stability of
ferromagnetism and although its strength is smaller than the Mn-Mn
one, it is enough to get a Curie temperature as high as 900 K
within MFA and 698 K within RPA. Thus both NiCrSi and NiMnSi are
very promising for realistic applications since their $T_C$ exceed
by far the room temperature of $\sim$300 K.

\begin{table*}
\centering \caption{ (Color online)Upper panel: FSKKR calculated
atomic spin moments for the NiCrSi(001) and NiMnSi(001) surfaces
for both possible terminations. We present the moments for the
atoms in the first two surface layers and in the last column their
sum. Lower panel: for each one of the atoms in the upper panel we
present the ratio of the majority-spin DOS at the Fermi level
($N^\uparrow$) vs the minority-spin DOS at the Fermi level
($N^\downarrow$). In the last column we present the
spin-polarization at the Fermi level calculated as
$\frac{N^\uparrow}{N^\uparrow + N^\downarrow}$ taking into account
both the surface and subsurface layers, $P$. \label{table3}}
\begin{tabular}{r|c|c|c|c|c}\hline

\multicolumn{6}{c}{NiMnSi surfaces} \\
& $m^{Ni}$ & $m^{Mn}$ & $m^{Si}$ & m$^{Void}$ & $m^{total}$  \\

(001)Ni &  0.455 & 3.166&   -0.116 & 0.037 &3.542 \\

(001)MnSi&   0.197  & 3.415 &  -0.213 & 0.013 &3.412  \\

\multicolumn{6}{c}{NiCrSi surfaces} \\
& $m^{Ni}$ & $m^{Cr}$ & $m^{Si}$ & m$^{Void}$ & $m^{total}$  \\
(001)Ni   & 0.287  &2.220 &-0.089 & 0.051& 2.439   \\

(001)CrSi  & 0.091 & 2.740 &  -0.180 &0.031 & 2.682 \\
\hline \hline
\multicolumn{6}{c}{NiMnSi surfaces} \\
& Ni ($N^\uparrow / N^\downarrow$) & Mn ($N^\uparrow /
N^\downarrow$) & Si ($N^\uparrow / N^\downarrow$) & Void
($N^\uparrow / N^\downarrow$) &
$P$ ($\frac{N^\uparrow}{N^\uparrow + N^\downarrow}$) \\

 (001)Ni      &  0.313/0.453  &0.412/0.288&   0.131/0.069
&0.045/0.049 & 51\% \\

(001)MnSi  & 0.242/0.058 & 0.235/0.191   & 0.116/0.051
& 0.043/0.013 & 67\% \\

\multicolumn{6}{c}{NiCrSi surfaces} \\
& Ni ($N^\uparrow / N^\downarrow$) & Cr ($N^\uparrow /
N^\downarrow$) & Si ($N^\uparrow / N^\downarrow$) & Void
($N^\uparrow / N^\downarrow$) &
$P$ ($\frac{N^\uparrow}{N^\uparrow + N^\downarrow}$)\\

(001)Ni      &  0.786/0.374& 2.045/0.237&   0.074/0.024
&0.082/0.043 &82\% \\

(001)CrSi&   0.460/0.063 & 1.263/0.155 & 0.083/0.065 & 0.052/0.015
&  86\% \\ \hline
\end{tabular}
\end{table*}

\section{Surfaces and Interfaces\label{sec5}}

\subsection{(001) Surfaces}

Although surfaces are not as interesting as interfaces we will
start our discussion from the (001) surfaces to give a basis for
our discussion in the case of interfaces. We have chosen this
surface orientation since this is the most usual one encountered
in experiments. We have to note first that we have considered only
abrupt surfaces and have not taken into account possible
reconstructions. When we open the (001) surfaces in NiCrSi and
NiMnSi we have two possible terminations. The surface layer can be
either a Ni-Void one with a Cr(Mn)Si as subsurface layer, or vice
versa. For the completeness of our results we have considered both
terminations and have used a slab of 15 layers embedded in vacuum
which has been shown to be adequate for the half-metallic Heusler
alloys (see Ref. \onlinecite{GalaSurf2}). We have also to note
here that the Ni-terminated surfaces contain also a vacant site
and are susceptible to show large reconstructions with respect to
the CrSi- and MnSi-terminated NiCrSi(001) and NiMnSi(001)
surfaces, respectively.\cite{Jenkins} Thus we will concentrate
mainly on the latter surfaces.

\begin{figure}
\centering
\includegraphics[scale=0.43]{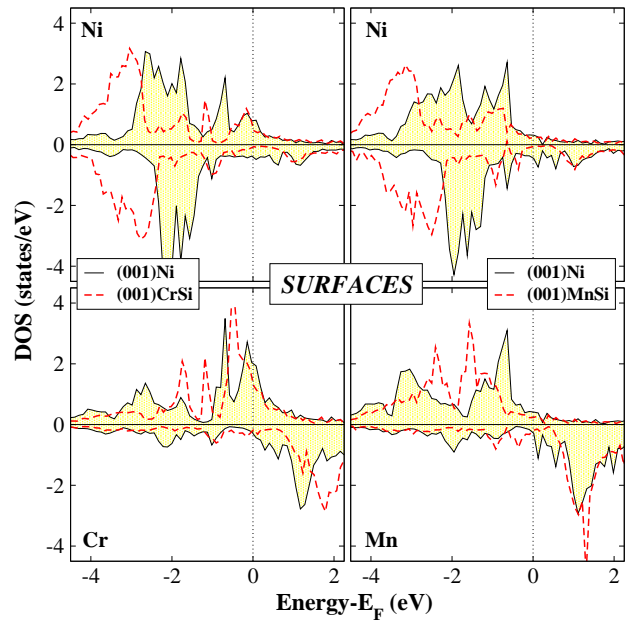}
\caption{ (Color online)FSKKR-calculated atom-resolved DOS for the
Ni and Cr(Mn) atoms sitting at the first two surface layers for
the NiCrSi(001) (left panel) and NiMnSi(001) (right panel)
surfaces. Note that e.g. there are two different terminations for
the NiCrSi(001) surface: a Ni surface layer with a CrSi subsurface
layer or a CrSi surface layer with a Ni subsurface layer.}
\label{fig8}
\end{figure}

In Fig. \ref{fig8} we present the DOS of the transition metal
atoms at the surface and subsurface layers for both NiCrSi and
NiMnSi alloys. It is obvious that in all cases the surfaces are
not half-metallic. The bonds which are broken at the surface
change the energy position of the states and the gap vanishes
similarly to what happens to semiconductors where surface states
make their surfaces metallic in most cases. More information we
can get from Table \ref{table3} where we present the spin moments
of the atoms at the surface and subsurface layers (which are
significantly enhanced with respect to the bulk due to the missing
hybrization) and also the number of majority ($N^\uparrow$) and
minority ($N^\downarrow$) electronic states at the Fermi level for
these atoms and the spin-polarization, $P$, defined as
$\frac{N^\uparrow}{N^\uparrow +N^\downarrow}$ taking into account
both the surface and subsurface layers.

We will start our discussion from the CrSi- and MnSi-terminated
surfaces. In the case of the CrSi-terminated surfaces the DOS with
the exception of the gap area is very similar to the bulk
calculations. The Ni atom in the subsurface layer presents
practically a half-ferromagnetic character with an almost zero
spin-down DOS, while for the bulk NiCrSi there is an absolute gap.
The Cr and Si atoms in the surface layer show more pronounced
differences with respect to the bulk, and within the gap  there is
a very small Cr-$d$ DOS. These states are strongly localized at
the surface layer as at the subsurface layer there is practically
no states inside the gap. When we open the CrSi(001) surface, the
Cr atoms at the surface layer loose two out of their four nearest
Ni neighbors.  This breaking of symmetry at the surface pushes the
Cr majority-spin states deeper in energy to account for the
missing hybridization and the charge accommodated in these bonds
is now accommodated by the Cr majority-spin states leading to a
considerable increase of the Cr atomic spin moment as it can be
seen also in Table \ref{table3}. Moreover the splitting between
the unoccupied Cr states above $E_F$ and the center of the
occupied Cr states decreases and at $E_F$ a surface state appears.
This behavior of the CrSi-terminated NiCrSi(001) surface is in
agreement with previous calculations on the MnSb-terminated
NiMnSb(001) surface.\cite{GalaSurf2} Similar is the situation also
for the MnSi-terminated NiMnSi(001) surface. Mn atoms act like the
Cr ones and accommodate most of the extra-charge due to the
broken-bonds in the majority-spin band. But Mn atoms  have one
valence electron more than Cr atoms which in the bulk is already
accommodated in the majority band and thus for Mn atoms the energy
cost to accommodate extra charge in the majority band is larger
than for the Cr atoms. As a result part of the extra-charge due to
the broken bonds is accommodated also in the minority-spin band
leading to a smaller increase of the spin moment with respect to
the Cr atoms and a more intense surface state within the gap. Mn
atoms at the surface layer show an increase of their spin moment
of about 0.6 $\mu_B$ with respect to the bulk reaching a value of
$\sim$3.4 $\mu_B$, while in the case of Cr atoms the increase of
the spin moment with respect to the bulk is 0.9 $\mu_B$.

In the case of the Ni terminated NiMnSi(001) surface, the Ni atoms
at the surface layer show a largely different shape of the DOS
with respect to the Ni atom at the subsurface layer in the case of
the MnSi-terminated surface and the bands are  located higher in
energy with respect to the MnSi(001) termination. The Mn atoms at
the subsurface layer show DOS very similar to the bulk case and
the spin magnetic moment is $\sim$3.17 $\mu_B$ very close to the
bulk value of $\sim$2.84 $\mu_B$. In this case it is clear that
there is a minority surface state located exactly at the Fermi
level since a very narrow peak centered at the Fermi energy
appears in the minority-spin DOS. This surface state survives also
for the Mn atoms at the subsurface layer. Similar arguments are
also valid for the Ni-terminated NiCrSi(001) surface. Cr atoms
behave as in the bulk with slightly larger spin moments. The spin
magnetic moment of the Ni atoms at the surface layer is
considerable larger than the value for the bulk (0.28$\mu_B$ for
the Ni surface atom in NiCrSi with respect to 0.17 $\mu_B$ for the
Ni atom in the bulk NiCrSi, and 0.46$\mu_B$ for the Ni surface
atom in NiMnSi with respect to 0.23 $\mu_B$ for the Ni atom in the
bulk NiMnSi) The surface state in the case of the Ni-terminated
NiCrSi(001) surface is more broad than in the case of the NiMnSi
alloy and cannot be distinguished from the rest of the DOS.

For applications more important is the spin-polarization at the
Fermi level. We present in the lower panel of Table \ref{table3}
the number of states at the Fermi level for each atom and for both
spin directions. We calculate the spin polarization as the number
of majority DOS divided by the total number of electronic states
at the Fermi level. In the case of Ni-terminated NiMnSi-surfaces
the number of states for the two spin directions is equal due to
the surface states and the spin-polarization is around 50\%\ while
for the MnSi-terminated surface the surface state is less intense
and the spin polarization reaches 67\% . In the case of NiCrSi
surface Cr atoms show a very large majority-spin DOS at the Fermi
level with respect to the Mn atoms since the Fermi level crosses
the $e_g$ states. This washes out the weight of the surface states
and NiCrSi surfaces show very large values of the spin
polarization: 82\%\ for the Ni-terminated surface and 86\%\ for
the CrSi-terminated surface. These results do not mean that NiCrSi
is more suitable for applications since the electronic properties
change considerably at the interfaces with semiconductors.

\begin{table*}
\centering \caption{ (Color online)Similar to the lower panel of
Table \ref{table3} for all studied interfaces between the
half-metals NiCrSi and NiMnSi and the GaP, ZnS and Si
semiconductors, taking into account all possible interface
structures. The spin-polarization ratios at the Fermi level have
been calculated as $\frac{N^\uparrow}{N^\uparrow + N^\downarrow}$
taking into account only the two first interface layers of the
half-metal, $P_1$, and the first two interface layers of the
half-metal and the first two interface layers of the
semiconductor, $P_2$.\label{table4}}
\begin{tabular}{r|c|c|r|c|c}\hline

\multicolumn{3}{c|}{NiMnSi/GaP Interface} &       \multicolumn{3}{c}{NiCrSi/GaP Interface} \\
& $P_1$ & $P_2$  &  & $P_1$ & $P_2$   \\
 Ni/Ga   & 0.963/0.567 (63\%)&
1.256/0.800 (61\%) & Ni/Ga & 2.538/0.569 (82\%) &2.771/0.835
(77\%)
\\
Ni/P   & 0.884/0.294 (75\%)
&1.127/0.462 (71\%) & Ni/P &3.212/0.225 (93\%)& 3.552/0.381 (90\%) \\
 MnSi/Ga& 0.905/0.810 (53\%)&
1.156/1.159 (50\%) & CrSi/Ga & 1.368/0.518 (73\%) &1.673/0.766 (69\%) \\
 MnSi/P  & 0.904/1.952 (32\%)&
1.222/2.575 (32\%) & CrSi/P& 1.561/0.768 (67\%) &1.813/1.037 (64\%)\\
\hline

\multicolumn{3}{c|}{NiMnSi/ZnS Interface}& \multicolumn{3}{c}{NiCrSi/ZnS Interface} \\
& $P_1$ & $P_2$ & & $P_1$ & $P_2$ \\
Ni/Zn &  0.868/0.538 (62\%)& 1.299/0.983 (57\%)&Ni/Zn &
2.244/0.337 (87\%)&
2.773/0.690 (80\%)\\
Ni/S & 0.796/1.100 (42\%) & 1.398/1.729 (45\%)&Ni/S    &
2.077/0.878 (70\%)&
2.747/1.433 (66\%)\\
MnSi/Zn &0.984/1.117 (47\%)& 1.522/1.724 (47\%)&CrSi/Zn  &
1.563/1.039 (60\%)&
2.010/1.706 (54\%)\\
MnSi/S & 0.761/1.005 (43\%)& 1.146/1.372 (46\%)& CrSi/S    & 1.903/1.037 (65\%)& 2.309/1.476 (61\%)\\
\hline

\multicolumn{3}{c|}{NiMnSi/Si Interface}   &    \multicolumn{3}{c}{NiCrSi/Si Interface}\\
& $P_1$ & $P_2$          &                     & $P_1$ & $P_2$  \\
Ni/Si & 0.993/0.282 (78\%) & 1.254/0.456 (73\%)     &
  Ni/Si & 3.016/0.163 (95\%) & 3.469/0.277 (93\%)\\
MnSi/Si  & 0.848/1.259 (40\%) & 1.129/1.854 (38\%) & CrSi/Si
&1.613/0.866 (65\%) & 1.873/1.140 (62\%) \\ \hline

\end{tabular}
\end{table*}

\subsection{Interfaces}

We will continue our study with the interfaces between the
half-metal and the semiconductors. Such structures are assumed to
present accurately the realistic situation since in most devices
half-metals are employed to inject current into a semiconductor.
We have assumed that the stacking direction is the (001) and used
a slab with 15 layers of the half-metal and 9 layers of the
semiconductor in order to have two equivalent interfaces as in
Ref. \onlinecite{GalaInter}. Moreover we have searched for
semiconductors crystallizing in the zinc-blende structure which is
compatible with the lattice structure of the Heusler
alloys\cite{Landolt} and with a lattice constant close to the
5.4\AA\ which is the equilibrium lattice parameter of both NiCrSi
and NiMnSi. Such semiconductors are the binary III-V semiconductor
GaP, the II-VI semiconductor ZnS and finally the IV semiconductor
Si. We have considered all possible interfaces and in Table
\ref{table4} we resume our results by presenting the
spin-polarization at the Fermi level taking into account the first
two half-metallic interface layers, $P_1$, and both the first two
half-metallic and the first two semiconducting layers, $P_2$. Note
here that contrary to previous studies we define the
spin-polarization $P$ as the ratio between the number of majority
(spin-up) states and the total number of states at the Fermi level
and thus $P$ represent directly the proportion of the majority
electrons at the Fermi level. There are two general remarks in
this table. First, when we consider also the semiconductor layers
the spin-polarization decreases slightly by a few percent in all
cases. The interface half-metallic layers polarize the
semiconducting layers at their vicinity and the first few
semiconducting layers become metallic. But the induced DOS of the
semiconductor within the gap is similar for both spin-directions
decreasing the spin-polarization. Second, the interfaces between
NiMnSi and the three semiconductors show considerably smaller
spin-polarization than the corresponding interfaces created by
NiCrSi. This is expected since, as for the surfaces, the Cr atoms
have a very high majority DOS at the Fermi level with respect to
the Mn atoms and they minimize the effect of interface states.
Thus in the following we will concentrate on the interfaces
between NiCrSi and the three different semiconductors.

\begin{figure}
\centering
\includegraphics[scale=0.43]{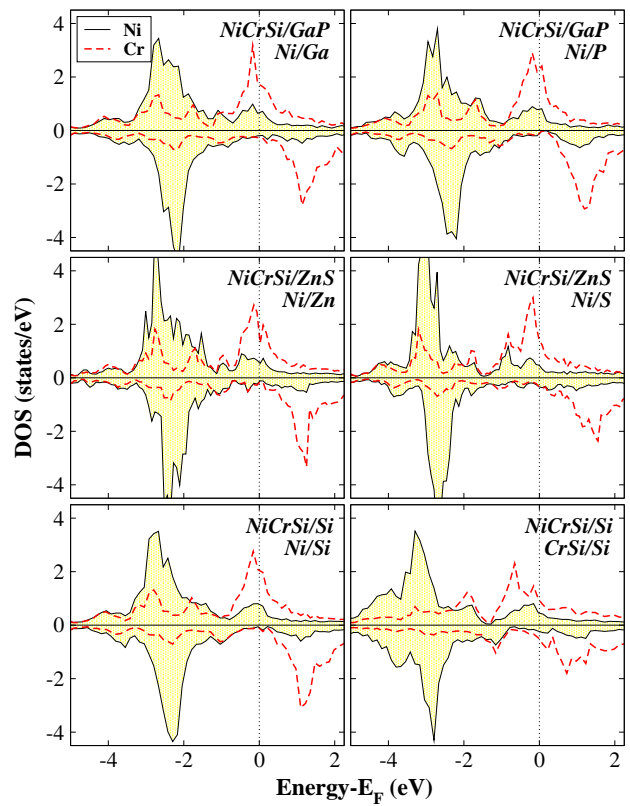}
\caption{ (Color online)FSKKR-calculated atom-resolved DOS for the
Ni and Cr atoms at the interface for the following cases: (i)
upper left panel: Ni/Ga structure for the NiCrSi/GaP interface,
(ii) upper right panel: Ni/P structure for the NiCrSi/GaP
interface, (iii) middle left panel: Ni/Zn structure for the
NiCrSi/ZnS interface, (iv) middle right panel: Ni/S structure for
the NiCrSi/ZnS interface, (v) lower left panel: Ni/Si structure
for the NiCrSi/Si interface, and (vi) lower right panel: CrSi/Si
structure for the NiCrSi/Si interface.}
 \label{fig9}
\end{figure}

\begin{table}
\centering \caption{ (Color online)FSKKR calculated atomic spin
moments for the two layers of the NiCrSi half-metal and of the two
first layers of the GaP, ZnS or Si semiconductors at the
interface. $Void$ denotes the empty site in the same layer with
the Ni atom. $Vc1$ denotes the empty site in the same layer with
Ga(Zn) and $Vc2$ in the same layer with the P(S) atoms. In the
case of the interfaces with Si, we denote with Si1 the Si atoms at
the interface layer and with Si2 the Si atoms at the subinterface
layer. \label{table5}}
\begin{tabular}{l|r|r|r|r|r|r|r|r}\hline

\multicolumn{9}{c}{NiCrSi/GaP Interface} \\
& \multicolumn{4}{c|}{Half-Metal}&
\multicolumn{4}{c}{Semiconductor}\\
& $m^{Ni}$ & $m^{Cr}$ & $m^{Si}$ & m$^{Void}$ & m$^{Ga}$ & m$^{Vc1}$ & m$^{P}$& m$^{Vc2}$\\

Ni/Ga  &  0.123& 2.050& -0.077&  0.046 & -0.008 &0.031& 0.040&
-0.003  \\

Ni/P& 0.132&  2.141& -0.080&  0.043 &  0.005 &0.002& -0.031
&0.027\\

CrSi/Ga& 0.196& 1.873& -0.041  & 0.037& -0.007
&0.053 &0.045  &  0.005  \\

CrSi/P& 0.248&  2.242& -0.015  &0.046 &  0.031 &0.019 &0.024
&0.077\\
\hline

\multicolumn{9}{c}{NiCrSi/ZnS Interface}\\
& \multicolumn{4}{c|}{Half-Metal} &
\multicolumn{4}{c}{Semiconductor} \\
& $m^{Ni}$ & $m^{Cr}$ & $m^{Si}$ & m$^{Void}$ & m$^{Zn}$ & m$^{Vc1}$ & m$^{S}$& m$^{Vc2}$\\

Ni/Zn &0.118 &1.955& -0.087  &0.039 & -0.012   &    0.025 &-0.001&
-0.005 \\

Ni/S &0.256&  2.400 &-0.064&  0.057 & 0.018& 0.004 &0.056& 0.028\\

CrSi/Zn &0.177&  2.087& -0.067 & 0.052& 0.022 &0.035& -0.011&
-0.028\\

CrSi/S &0.168 & 1.527 &-0.035 &0.037 &0.004 &0.004 &0.032 &0.025\\
\hline

\multicolumn{9}{c}{NiCrSi/Si Interface}\\
& \multicolumn{4}{c|}{Half-Metal} &
\multicolumn{4}{c}{Semiconductor} \\
& $m^{Ni}$ & $m^{Cr}$ & $m^{Si}$ & m$^{Void}$ & m$^{Si1}$ &
m$^{Vc1}$ & m$^{Si2}$
& m$^{Vc2}$  \\

Ni/Si &0.132&  2.080 &-0.081&  0.042 & -0.014& 0.030 &0.025& 0.001
\\

CrSi/Si &0.198 & 1.939 &-0.025 &0.038 &0.005 &0.066 &0.049
&0.013\\ \hline
\end{tabular}
\end{table}

When the contact at the interface is created by the Ni atomic
layer and not the CrSi layer the obtained spin-polarization is
much higher and can exceed 90\%\ as in the case of Ni/P and Ni/Si
contacts. The different behavior between the Ni and CrSi interface
layers can be easily understood in terms of symmetry and
hybridization. Ni interface atoms have a spin moment close to
their bulk value as shown in Table \ref{table5} where we present
the atomic spin magnetic moments for all atoms at the interface.
In the bulk case Ni has 4 Cr and 4 Si atoms as first neighbors. On
the Ni-terminated (001) surface the Ni atom loses half of its
first neighbors. But if an interface with a semiconductor is
formed, the nickels two lost Si neighbors are replaced by two $sp$
atoms presenting similar electronic structure to the Si ones,
although they have in general different number of valence
electrons, and the situation is much closer to the bulk case than
when we have CrSi as an interface layer. Now the Si $p$ bands at
lower energy are not destroyed since the $sp$ atoms have a
behavior similar to Si and still they accommodate transition metal
$d$ electrons. Thus the only change in the Ni DOS comes from the
missing two Cr neighboring atoms. The DOS of the Ni atom at the
interface layer presented in Fig.~\ref{fig9} for  several Ni/$sp$
contacts is clearly very close to the bulk case. Moreover the Cr
atoms at the subinterface layer are also almost bulk like showing
with the exception of the gap area a DOS close to the bulk one for
all cases of Ni/$sp$ contacts. Between the different cases
presented in Fig. \ref{fig9} -Ni/Ga and Ni/P contacts in the case
of NiCrSi/GaP interfaces, Ni/Zn and Ni/S contacts in the case of
NiCrSi/ZnS interfaces and Ni/Si contact in the case of NiCrSi/Si
interface- the DOS of the Ni interface atoms and Cr subinterface
atoms only marginally change. The small variations at the gap area
arise from the different hybridization between the $t_{2g}$
$d$-orbitals of Ni and the $p$-orbitals of the $sp$ atom which
transform with the same representation and thus can couple. When
the $sp$ atom is Si the situation is closer to the bulk NiCrSi and
we get a $P_2$ spin polarization as high as 93\%. When the $sp$
atom is P which has one electron more than Si the $P_2$ drops
slightly to 90\%. As the difference between the electronic
structure of Si and the $sp$ atoms becomes more important the
hybridization effects become more pronounced inducing a larger DOS
in the minority-spin band and thus leading to smaller values of
$P_2$.

In the case when the contact is made up by the CrSi layer the Cr
atoms have an immediate environment very different from the bulk
NiCrSi and thus their electronic properties are considerably
altered with respect to the bulk case. In Fig. \ref{fig9} we
present the DOS of the Cr atom at the interface layer and the Ni
atom at the subinterface layer in the case of a CrSi/Si contact.
The majority-spin bands are broader with respect to the Ni/Si
contact and they are pushed deeper in energy. This  attracts some
of the minority unoccupied Cr states also deeper in energy and now
the Fermi level crosses the antibonding minority $d$-bands leading
to a smaller spin-polarization at the Fermi level. While the
CrSi-based contacts keep a degree of polarization close to 60\%\
due to the large intensity of the Cr majority-spin DOS at the
Fermi level, in the case of the MnSi-based contacts the situation
is even worse and $P_2$ becomes smaller than 50\%\ meaning that at
the Fermi level the spin-down states are more than the spin-up
states.

Finally we should also briefly discuss the spin magnetic moments
presented in Table \ref{table5} for the case of NiCrSi/(GaP or ZnS
or Si) interfaces. The atoms in the semiconductor spacer become
magnetic due to the polarization from the transition-metal atoms
in the half-metallic spacer but their spin moments are very small
(less than 0.01$\mu_B$ in all cases). Ni and Cr  atoms in all
cases present spin moments close to the bulk values and there is
no general trend associating the spin magnetic moments to the
spin-polarization of the interfaces. Thus for each case we have to
study the DOS at the interface to get valid conclusions about the
behavior of the half-metallicity.

 We should also note here that we have neglected relaxation
 effects at the interface. Relaxation effects modify the spin-polarization
 at the Fermi level since it leads to
increased hybridization and charge-transfer. It was shown in the
case of a Ni/P contact in the case of the NiMnSb/GaP interface
that the Ni-P interlayer distance was reduced by 18\%, while the
neighboring interlayer distances where expanded by 5-7\% as
compared to the ideal bulk values leading to a decrease of the
spin-polarization.\cite{GalaInter} Nevertheless, these effects did
not destroy the gap but shifted the local DOS slightly deeper in
energy and a suitable doping at the interface restored the
spin-polarization. Moreover in the case where the
spin-polarization for the unrelaxed surface is as high as 90\%\ as
in some of the cases presented above, the effect of relaxation
will be much weaker.

\begin{figure*}
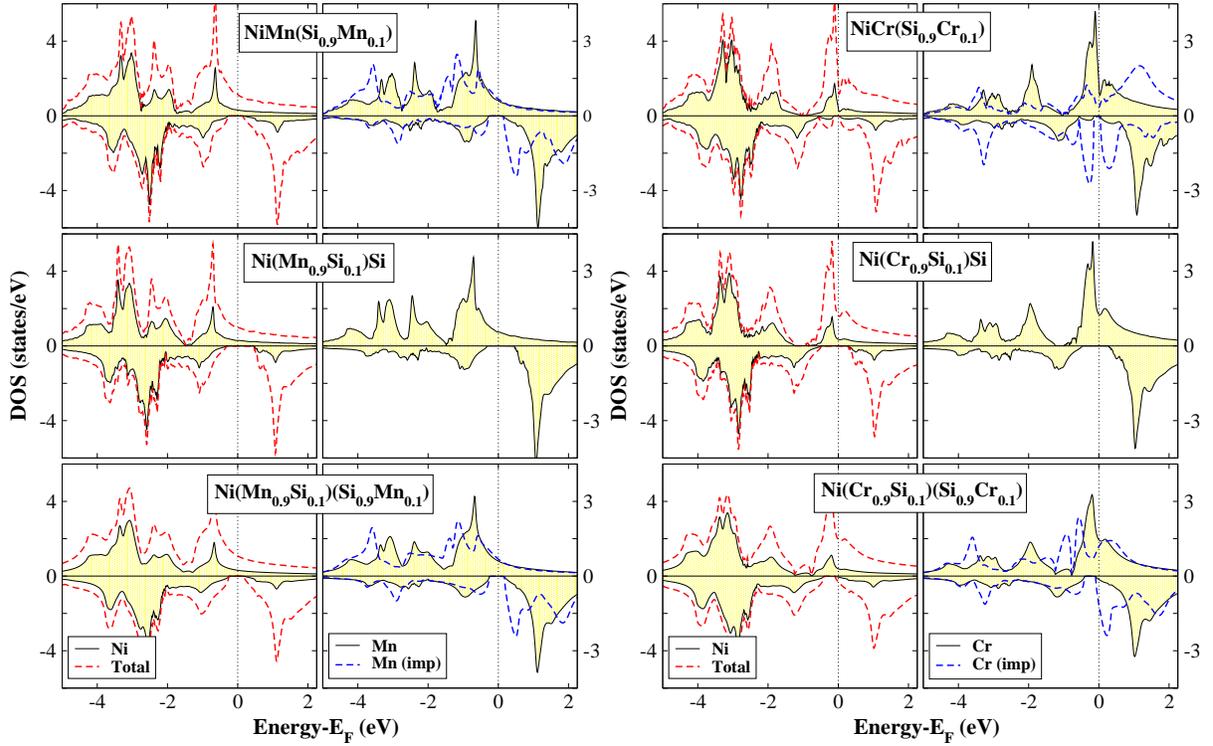

\centering
\includegraphics[scale=0.4]{fig10a.eps}
\includegraphics[scale=0.4]{fig10b.eps}
\caption{ (Color online)Left panel: FPLO-calculated total and atom
resolved DOS for three distinct doping cases in NiMnSi which
conserve the half-metallicity. Right panel: similar DOS for
NiCrSi. Atomic DOS have been scaled to one atom.}
 \label{fig10}
\end{figure*}

\begin{table}
\centering \caption{ (Color online)FPLO calculated width of the
minority-spin gap in eV for the cases presented in figure
\ref{fig10} as a function of the disorder concentration $x$. With
``--'' we denote the cases where the gap persists but the
half-metallicity is lost since the Fermi level is above the gap.
\label{table6}}
\begin{tabular}{l|c|c} \hline

&\multicolumn{2}{c}{NiY(Si$_{1-x}$Y$_x$)}\\
$x$ & Y=Mn & Y=Cr \\
 0      &   1.055 & 1.071 \\ 0.05 &
0.584 & 0.298\\ 0.1  & 0.437 & 0.176  \\0.2& 0.206 & -- \\ \hline

&\multicolumn{2}{c}{Ni(Y$_{1-x}$Si$_x$)Si}\\
$x$ & Y=Mn & Y=Cr \\
0 & 1.055 & 1.071 \\ 0.05 &0.878 & 0.878\\0.1 & 0.777 & 0.773 \\ 0.2 & 0.626& 0.613\\
\hline

&\multicolumn{2}{c}{Ni(Y$_{1-x}$Si$_x$)(Si$_{1-x}$Y$_x$)}\\
$x$ &Y=Mn & Y=Cr \\
0 & 1.055 & 1.071 \\ 0.05 & 0.567 & 0.483 \\ 0.1 & 0.424 & --\\
0.2 & 0.261& --\\ \hline

\end{tabular}
\end{table}

\section{Disorder and defects \label{sec6}}
Our final section is devoted to the effect of defects. Defects in
large concentrations can induce new bands within the minority-spin
gap and completely destroy half-metallicity. But even in low
concentrations they can couple to interface states and the latter
ones become conducting destroying the spin polarization during
transport experiments through a half-metal/semiconductor
interface. To perform our study we have employed the FPLO method
within the CPA approximation to account for disorder in a random
way. We have studied two types of defects (i) the creations of
Mn(Cr) and Si antisites and Mn(Cr)-Si swaps which keep the
half-metallic character, and (ii) the creation of Ni antisites and
Ni-Mn(Cr) atomic swaps which induce states within the gap
destroying the gap.

\subsection{Defects conserving the half-metallicity}

As we have mentioned just above the creation of Cr(Mn) and Si
antisites keeps the half-metallic character, as-well-as the
Cr(Mn)-Si atomic swaps. The reason is pretty simple. If we examine
the band-structure of the minority-spin electrons in Fig.
\ref{fig5} the states just below the gap are the $t_{2g}$
$d$-hybrids which are mainly located at the Mn(Cr) sites. Above
the gap we have the antibonding $e_g$ $d$-states and then the
antibonding $t_{2g}$ $d$-hybrids. The $t_{2g}$ states transform
using the same representation as the $p$ states of the Si atom.
The sites occupied by the Si and Cr(Mn) atoms in the perfect
compounds have the same symmetry, with four Ni and four Void sites
as first neighbors, rotated by 90$^o$. Thus the equivalence
between the $t_{2g}$ and $p$ states is obvious. When Cr(Mn) atoms
migrate to Si sites or Si atoms to Cr(Mn)sites, it results to a
broadening of the bonding $t_{2g}$ hybrids just below the gap.
Also the antibonding $t_{2g}$ band above the gap broadens and it
overlaps with the antibonding $e_g$ states. These two simultaneous
phenomena lead to a shrinking of the gap which as we will discuss
below persists for a moderate percentage of disorder. Of course
above a critical value of disorder the gap vanishes. In the case
of Cr(Mn)-Si atomic swaps, the phenomenon discussed above is more
intense and the gap vanishes for a smaller degree of disorder.

In Fig. \ref{fig10} we present the total, Ni, and Cr(Mn) DOS for
the case of 10\%\ Cr(Mn) antisites [NiCr(Si$_{0.9}$Cr$_{0.1}$) and
NiMn(Si$_{0.9}$Mn$_{0.1}$) alloys], the case of 10\%\ Si antisites
[Ni(Cr$_{0.9}$Si$_{0.1}$Si) and Ni(Mn$_{0.9}$Si$_{0.1}$)Si], and
10\%\ atomic swaps [Ni(Cr$_{0.9}$Si$_{0.1}$)(Si$_{0.9}$Cr$_{0.1}$)
and  Ni(Mn$_{0.9}$Si$_{0.1}$)(Si$_{0.9}$Mn$_{0.1}$)]. We also
present for the first and third case the DOS of the Cr(Mn)
impurity atoms at the antisites. First remark is that Ni DOS  is
similar for all three compounds and similar to the Ni DOS in the
perfect compounds. The same is true for the Cr and Mn atoms at the
perfect sites. Thus the antisites and atomic swaps have only
marginal effect on the properties of the transition-metal atoms at
perfect sites. Concerning now the impurity Cr and Mn atoms sitting
at perfect Si sites, the hybridization between the Cr(Mn) $t_{2g}$
states and $p$ states leads to important differences with respect
to the Cr and Mn atoms sitting at the perfect sites. The
minority-spin  impurity states move lower in energy but in the
case of the Mn impurities the large exchange splitting between the
majority and minority $d$-states keeps these states well above the
gap. In the case of the Cr impurities the exchange splitting is
significantly lower than for the Mn impurities and the Cr
unoccupied minority $e_g$ states are crossed by the Fermi level in
the case of atomic swaps (lower panel). In the case of Cr
antisites the antibonding $e_g$ states move even lower in energy
and the Fermi level falls within a tiny gap separating the
antibonding $e_g$ and $t_{2g}$ states and the Cr impurity atoms
have an almost vanishing spin magnetic moment. (We do not present
the spin moments of the other atoms since they are very close to
the values for the perfect compound.)

\begin{figure}
\centering
\includegraphics[scale=0.43]{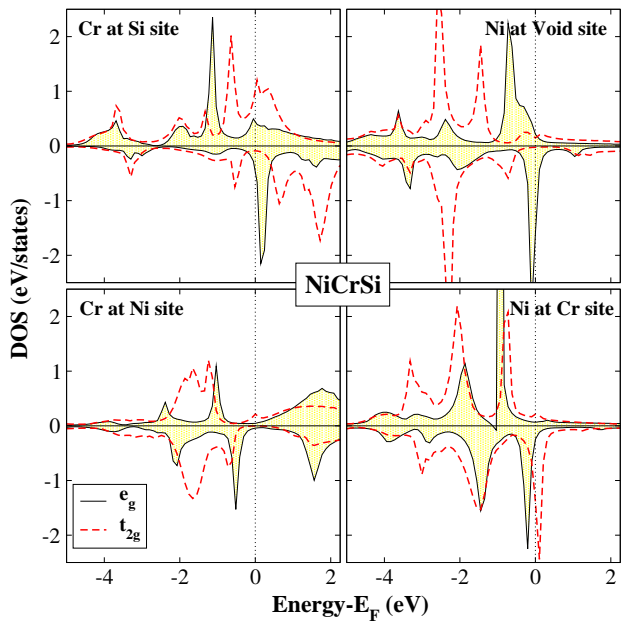}
\caption{ (Color online)$d$-resolved atomic DOS for several cases
of single impurities  in NiCrSi compound calculated using the
FSKKR method. The DOS has been decomposed in the
double-degenerated $e_g$ and triple-degenerated $t_{2g}$
constituents.}
 \label{fig11}
\end{figure}

To confirm the character of the Cr impurity states we have also
employed FSKKR method and have made calculations for a single Cr
impurity in  NiCrSi and in Fig. \ref{fig11} we present the $d$
states of Cr antisites at a Si site projected on the $e_g$ and
$t_{2g}$ orbitals. For the impurity calculations we perform the
self-consistent calculations for NiCrSi and produce the Green
function for the bulk compound.\cite{imp1,imp2} Then we consider a
cluster of 65 atoms surrounding the Si site embedded in an
infinite NiCrSi bulk crystal and calculate the electronic
structure of the atoms in this cluster in real space considering a
Cr atom at the central Si site. (We have to note here that we have
neglected the relaxation of the atomic positions within the
cluster due to the impurity atom, which is very tedious
computationally, since we expect them to be small due to the high
symmetry of the crystal and an extensive study of impurities is
out of the scope of the present manuscript.) As we can deduce from
the DOS in Fig. \ref{fig4} the antibonding unoccupied minority
$d$-states have mainly their weight at the Cr site while the
bonding occupied minority states at the Ni site. The minority
$e_g$-states of the Cr-impurity are located lower in energy with
respect to the Cr atom in the perfect site. Moreover they are
located exactly above the Fermi level and are also very localized
in energy. The $t_{2g}$ minority states of the Cr-impurity atom
form a wider band which starts just at the right edge of the $e_g$
peak and goes very high in energy. In the case of atomic swaps the
$e_g$ states move even lower in energy forming this tiny gap. The
Cr-impurity atom polarizes slightly the neighboring atoms but the
effect very quickly vanishes inside the cluster. As we can deduce
from these results on a single impurity in a perfect bulk crystal,
which are in very good agreement with the results obtained for a
disordered crystal within the coherent potential approximation
(CPA) where each disordered site is occupied by a pseudoatom with
mixed Si-Cr properties, local-order effects and long-range effects
have the same influence on the electronic properties of NiCrSi
alloy. We have performed similar calculations for a single Mn
impurity atom sitting at a Si site in NiMnSi and the results were
similar to the case of a Cr antisite in NiCrSi but now the $e_g$
and $t_{2g}$ states in the minority-spin band where higher in
energy due to the larger exchange-splitting of the Mn occupied
majority and unoccupied minority $d$ states.

Finally, we should discuss the width of the gap which is the most
interesting property affected by the defects. In table
\ref{table6} we present the width of the gap as a function of the
concentration of defects for all three studied cases and for both
Cr- and Mn-based compounds. The more mild effect is the creation
of Si antisites (middle panel of the table) since
no-transition-metal impurities exist. The gap persists with large
width-values even for as much as 20\%\ of Si antisites (0.626 eV
for the Mn-based compound and 0.613 eV for the Cr-based alloy).
When we create atomic swaps (lower panel of the table)  the
gap-width shrinks with a double rate and for NiMnSi alloys it
drops to 0.216 eV for the case of 20\%\ atomic swaps. In the case
of NiCrSi the gap-width is half its initial value for 5\%\ of
atomic swaps (0.484 eV instead of 1.071 eV). For 10\%\ creation of
Cr-Si atomic swaps the gap persists but the Fermi level is above
the gap and the defected alloy is not half-metallic. As we have
mentioned above in the case of simple Mn(Cr) antisites the gap
shrinks very quickly similarly to what happens for the atomic
swaps but in the case of Cr antisites in NiCrSi we have to reach
20\%\ of Cr antisites to loose half-metallicity. Thus for moderate
values of antisites and atomic swaps (around 5-10\%) both NiCrSi
and NiMnSi remain half-metallic although the gap shrinks with
respect to the perfect alloys.

\begin{figure}
\centering
\includegraphics[scale=0.43]{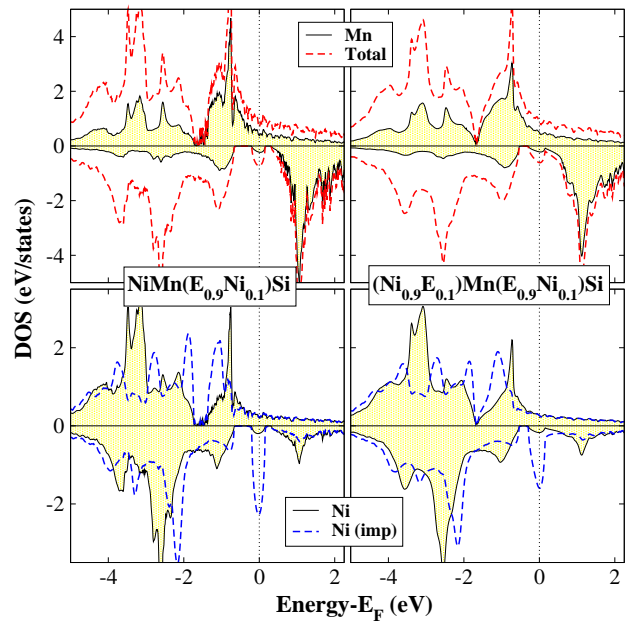}
\caption{ (Color online)Left panel: FPLO-calculated total and atom
resolved DOS for two distinct doping cases in NiMnSi which destroy
the half-metallicity.  Note that with E we denote the vacant site.
Atomic DOS have been scaled to one atom.}
 \label{fig12}
\end{figure}

\subsection{Defects destroying the half-metallicity}

We will proceed now with the case of defects destroying the gap
and firstly the creation of Ni antisites at the void site. We have
studied two cases (i) there is a surplus of Ni atoms at the void
site [NiCr(E$_{1-x}$Ni$_{x}$)Si and NiMn(E$_{1-x}$Ni$_{x}$)Si
compounds where E denotes the void site], and (ii) Ni atoms
migrate from the perfect Ni site to the void site conserving the
number of Ni atoms [(Ni$_{1-x}$E$_{x}$)Cr(E$_{1-x}$Ni$_{x}$)Si and
(Ni$_{1-x}$E$_{x}$)Mn(E$_{1-x}$Ni$_{x}$)Si compounds]. We present
in Fig. \ref{fig12} the DOS for the Mn-based compounds and for
$x=0.1$ for both kind of Ni antisites. The results are similar for
the Cr-based alloys and thus we do not present the DOS in this
case. We see than in all cases there is a clear peak instead of
the minority-spin gap. This is not present only for the Ni
impurity atom but also for the Ni and Mn atoms at the perfect
sites. In the case of Ni surplus the intensity of the peak is
larger and is more localized in energy. All other details of the
DOS are the same with the perfect NiMnSi compound and the Ni
impurity atom has a similar DOS with the Ni atoms at the perfect
sites since the Void and Ni sites have the same symmetry (four Mn
and four Si atoms as first neighbors). This peak is present also
in the case of the NiCrSi alloys and it has the same shape and
intensity. This gives the hint that this peak comes from only one
band. To investigate it further we show in Fig. \ref{fig11} also
the case of a single Ni impurity at a Void site in NiCrSi with the
FSKKR impurity code (the details are the same with the case of Cr
at Si site in NiCrSi discussed above). Clearly the peak comes from
the $e_g$ electrons which form a very narrow band and the Fermi
level is located exactly at this peak. Contrary to the case of a
single Cr impurity at a Si site, this peak is present for all
neighboring atoms and survives practically for most of the atoms
of the cluster showing that the presence of Ni impurity atoms
provokes a shift of the minority unoccupied antibonding $e_g$ band
lower in energy and this band has its weight mainly on the Cr(Mn)
atoms. We have also examined the cases where $x$= 0.05 and 0.2 for
both NiCrSi and NiMnSi and for both kind of Ni antisites. Even for
$x$=0.05 the peak is present but now it is more narrow with larger
intensity, and as we increase the concentration of antisites the
peak starts to occupy a larger energy range and its intensity
becomes smaller. Thus we can safely conclude that Ni antisites at
the vacant site completely kill the spin-polarization.

\begin{figure}
\centering
\includegraphics[scale=0.43]{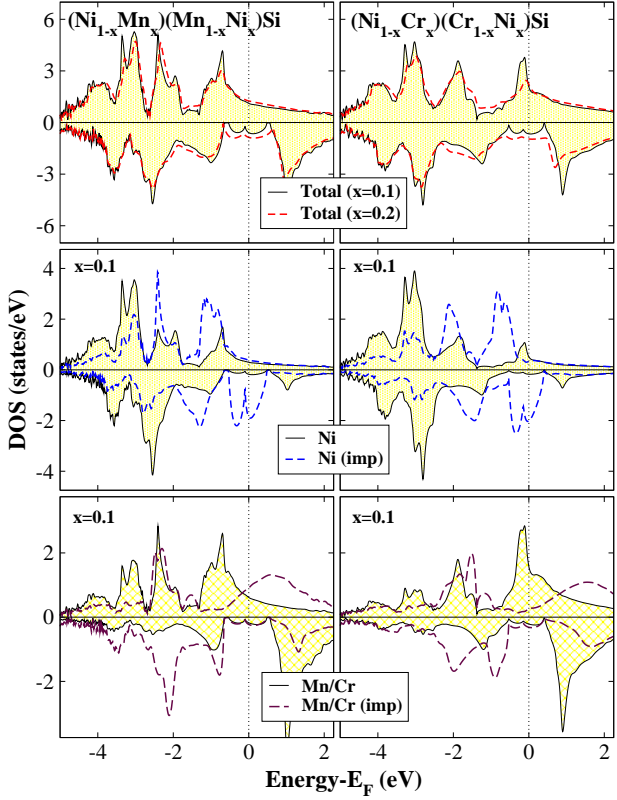}
\caption{ (Color online)FPLO-calculated total and atom resolved
DOS for the case of Mn-Ni atomic swaps in NiMnSi (left panel) and
Cr-Ni atomic swaps in NiCrSi (right panel). Atomic DOS have been
scaled to one atom.}
 \label{fig13}
\end{figure}

\begin{table}
\centering \caption{ (Color online)FPLO calculated atomic spin
moments in $\mu_B$ for the case of Mn-Ni and Cr-Ni atomic swaps in
NiMnSi and NiCrSi compounds, respectively, within the CPA
approximation. In the case of the perfect compounds ($x=0$) we
give in parenthesis the spin moment of a single impurity atom
calculated using the FSKKR method. \label{table7}}
\begin{tabular}{l|r|r|r|r|r|r} \hline

& \multicolumn{6}{c}{(Ni$_{1-x}$Mn$_x$)(Mn$_{1-x}$Ni$_x$)Si} \\

$x$ & $m^{Ni}$ & $m^{Ni(imp)}$ & $m^{Mn}$ & m$^{Mn(imp)}$ & m$^{Si}$ &  $m^{total}$ \\

0 & 0.207 &  (0.424) & 3.005 & (-1.451) &-0.212 & 3.000 \\

0.1& 0.180& 0.480& 2.849 & -1.496 & -0.125 & 2.509 \\

0.2& 0.186 & 0.302 & 2.674 &-1.358 & -0.096 & 1.988 \\ \hline

& \multicolumn{6}{c}{(Ni$_{1-x}$Cr$_x$)(Cr$_{1-x}$Ni$_x$)Si} \\

$x$ & $m^{Ni}$ & $m^{Ni(imp)}$ & $m^{Cr}$ & m$^{Cr(imp)}$ & m$^{Si}$ &  $m^{total}$ \\

0 & 0.127& (0.429) &2.020 & (-0.870) & -0.148 & 2.000 \\

0.1& 0.078& 0.409 & 1.652 & -0.943 & -0.081 & 1.433 \\

0.2& 0.054& 0.244&1.196 & -0.656& -0.052& 0.870 \\ \hline
\end{tabular}
\end{table}

Finally, we have investigated the case of Ni-Cr and Ni-Mn atomic
swaps. These defects have the peculiarity that Cr(Mn) and Ni atoms
move to sites of different symmetry with respect to their perfect
sites in the NiCrSi and NiMnSi alloys. In Fig. \ref{fig13} we
present for both compounds the total DOS for two different
concentrations of atomic swaps ($x$=0.1 and 0.2) and the
atom-resolved DOS for the first case. The Ni and Cr(Mn) atoms at
the perfect sites show a DOS similar to the perfect compounds
presented in Fig. \ref{fig4}. The impurity Ni atoms at the perfect
Cr(Mn) sites show a double-peak structure and the Fermi level
falls within this structure. When the concentration is increased
to 0.2, as it is obvious from the total DOS, the two peaks overlap
creating a wide-band. The Ni impurity atoms carry a very large
spin moment with respect to the spin moment of the Ni atoms at the
perfect Ni site as can be seen in Table \ref{table7} where we
present the spin magnetic moments for both Cr- and Mn-alloys and
for various concentrations of atomic swaps. As we increase the
concentration of atomic swaps the spin moments of both the Ni
impurity atoms and Ni atoms at the perfect sites decrease. The
changes are even more important for the Cr(Mn) atoms at the Ni
sites. Due to the different symmetry these atoms have a spin
moment antiparallel to the spin moment of the other transition
metal atoms reducing considerably the total spin moment and
leading to ferrimagnetic alloys. This behavior is present also in
the case of Cr and Mn impurity atoms at perfect Co sites in
Co$_2$CrAl(Si) and Co$_2$MnAl(Si) full-Heusler
alloys,\cite{SSC1,SSC2} where Cr(Mn) impurity atoms have spin
moments antiparallel to the spin magnetic moment of the other
transition metal atoms. But in these full-Heusler alloys the
half-metallic character is kept contrary to the compounds under
study here. This is due to the Ni impurity atoms and not to the
Cr(Mn) impurity atoms which have a very small DOS in the
minority-spin band which is obviously a very weak image of the
Ni-impurity DOS. To examine this behavior even further we have
plotted in Fig. \ref{fig11} the DOS for a single Cr impurity atom
at a perfect Ni site and for a single Ni impurity atom at a
perfect Cr site in NiCrSi. In the case of the Cr impurity atom the
gap persists and the spin moment is negative (in Table
\ref{table7} we present in parenthesis the spin moment of the
single impurity atoms) and close to the values calculated using
the FPLO-CPA code. The Ni impurity atoms at the Cr site show a
double peak structure at the Fermi level with the $e_g$ states
just below the gap and the Fermi level being pinned exactly at the
minority $t_{2g}$ peak. Thus due to symmetry reasons the majority
$d$-states of Ni-impurity atoms  are completely occupied leading
to large spin moments and the Fermi level falls within a double
peak-structure leading to a complete destruction of
half-metallicity. Thus any kind of Ni defects leads to loss of the
half-metallic character.

\section{Summary and conclusions \label{sec7}}

We have studied using both the full--potential nonorthogonal
local--orbital minimum--basis band structure scheme (FPLO) and the
full-potential screened Korringa-Kohn-Rostoker (FSKKR) electronic
structure methods the electronic, magnetic and gap-related
properties of the NiYSi compounds and have expanded our study also
to cover the case of surfaces, interfaces with semiconductors and
defects. When Y stands for V, the ferromagnetism is not very
stable due to the weak V-V interactions and these compounds are
not suitable for applications.  When Y stands for Cr or Mn, the
ferromagnetism is extremely stable leading to very high values of
the Curie temperature which is predicted to be $\sim$700 K for
NiCrSi and $\sim$1100 K for NiMnSi.

Both NiCrSi and NiMnSi are half-metallic at their equilibrium
lattice constant with large width of the minority-spin gap
($\sim$1 eV) and integer values of the total spin moment as
predicted by the Slater Pauling rule. The gap is created due to
the creation of bonding and antibonding $d$-hybrids in the
minority-spin band. The width of the gap is marginally affected
upon tetragonalization even when we expand or contract the lattice
by 5\%.

NiCrSi (001) surfaces present a high spin-polarization at the
Fermi level with respect to the NiMnSi alloy due to the large
intensity of the Cr majority-spin density of states (DOS) at the
Fermi level. In the case of interfaces with semiconductors with
similar lattice parameter (GaP, ZnS and Si), Ni-based contacts
show larger spin-polarization since Ni atoms at the interface
layer have a more bulk-like environment and in the case of NiCrSi
interfaces the high Cr-DOS leads to values of the
spin-polarization for the Ni-based contacts as high as 90\%.

Finally, we have shown that there are two kind of defects. The
defects and atomic swaps involving the Cr(Mn) and Si atoms lead to
a broadening of the bands and the gap is slowly shrinking and for
a critical value of the defects-concentration it disappears. Ni
defects on the other hand lead to a shift of the energy-localized
$e_g$ states within the gap and these impurity states completely
destroy the half-metallicity.

For realistic applications it seems that NiCrSi is more suitable
with respect to both NiMnSi studied here and the well-known and
widely studied NiMnSb Heusler alloy. Its Curie temperature is
smaller than the Mn-based alloys but it is still high-enough for
applications exceeding considerably the room temperature. Moreover
the large population of Cr majority-spin states at the Fermi level
ensures the high-spin polarization at the Fermi level which is
needed for the injection of current in semiconductors. Crucial for
the operation of devices based on Heusler alloys is the prevention
of the creation of Ni defects, since they induce impurity states
within the gap which can couple to interface states and completely
destroy the spin-polarization of the current injected into the
semiconductor.

\end{document}